\documentclass[sigconf,natbib=true]{acmart}
\AtBeginDocument{%
  }

\copyrightyear{2024}
\acmYear{2024}
\setcopyright{acmlicensed}\acmConference[WSDM '24]{Proceedings of the 17th ACM International Conference on Web Search and Data Mining}{March 4--8, 2024}{Merida, Mexico}
\acmBooktitle{Proceedings of the 17th ACM International Conference on Web Search and Data Mining (WSDM '24), March 4--8, 2024, Merida, Mexico}
\acmPrice{15.00}
\acmDOI{10.1145/3616855.3635817}
\acmISBN{979-8-4007-0371-3/24/03}

\usepackage{algorithm}
\usepackage{algorithmicx}
\usepackage{algpseudocode}

\usepackage{xcolor}
\usepackage{tikz}
\usepackage{multirow}
\usepackage{booktabs}
\usetikzlibrary{patterns}
\usepackage{pgfplots}
\usepackage{amsmath}
\usepackage{upgreek}
\usepackage{enumitem}
\newlist{longenum}{enumerate}{1}
\setlist[longenum,1]{leftmargin=4mm, label=\roman*)}

\pgfplotsset{compat=1.5.1}
\newenvironment{customlegend}[1][]{%
  \begingroup
  \csname pgfplots@init@cleared@structures\endcsname
  \pgfplotsset{#1}%
}{%
  \csname pgfplots@createlegend\endcsname
  \endgroup
}%
\def\addlegendimage{\csname pgfplots@addlegendimage\endcsname}
\usetikzlibrary{patterns}
\usetikzlibrary{pgfplots.groupplots}
\usetikzlibrary{
  pgfplots.colorbrewer,
}

\newcommand{\spec}{{\it spec.}}

\newcommand{\ie}{{\it i.e.}}
\newcommand{\eg}{{\it e.g.}}

\newcommand{\ours}{\textsc{\textsf{{MONET}}}}

\begin{document}

\title{MONET: Modality-Embracing Graph Convolutional Network and Target-Aware Attention for Multimedia Recommendation}

\author{Yungi Kim}
\authornote{Two first authors have contributed equally to this work.}
\affiliation{
	\institution{Hanyang University}
	\city{Seoul}
  	\country{Korea}
}
\email{gozj3319@hanyang.ac.kr}

\author{Taeri Kim}
\authornotemark[1]
\affiliation{
	\institution{Hanyang University}
	\city{Seoul}
  	\country{Korea}
}
\email{taerik@hanyang.ac.kr}

\author{Won-Yong Shin}
\affiliation{
	\institution{Yonsei University}
	\city{Seoul}
  	\country{Korea}
}
\email{wy.shin@yonsei.ac.kr}

\author{Sang-Wook Kim}
\authornote{Corresponding author.}
\affiliation{
	\institution{Hanyang University}
	\city{Seoul}
  	\country{Korea}
}
\email{wook@hanyang.ac.kr}



\begin{abstract}
In this paper, we focus on multimedia recommender systems using graph convolutional networks (GCNs) where the multimodal features as well as user--item interactions are employed together. 
Our study aims to exploit multimodal features more effectively in order to accurately capture users' preferences for items.
To this end, we point out following two limitations of existing GCN-based multimedia recommender systems: (L1) although multimodal features of interacted items by a user can reveal her preferences on items, existing methods utilize GCN designed to focus {\it only} on capturing collaborative signals, resulting in insufficient reflection of the multimodal features in the final user/item embeddings; (L2) although a user decides whether to prefer the target item by considering its multimodal features, existing methods represent her as only a single embedding {\it regardless of the target item's multimodal features} and then utilize her embedding to predict her preference for the target item.
To address the above issues, we propose a novel multimedia recommender system, named \ours, composed of following two core ideas: {\it modality-embracing} GCN (MeGCN) and {\it target-aware} attention. Through extensive experiments using four real-world datasets, we demonstrate i) the significant superiority of \ours~over seven state-of-the-art competitors (up to 30.32\% higher accuracy in terms of recall@20, compared to the best competitor) and ii) the effectiveness of the two core ideas in \ours.
All {\ours} codes are available at \url{https://github.com/Kimyungi/MONET}.
\end{abstract}

\maketitle

\section{Introduction}\label{sec:introduction}
To alleviate the {\it sparse nature} of user--item interactions (\eg, user's purchase history and click log) that reveal the user's preferences, many recommender systems tend to exploit side information such as {\it multimodal data} of items including textual and visual modalities~\cite{VBPR, JRL, VECF, CMF, DeepStyle, CDR, CTR, TopicMF, MMGCN, GRCN, LATTICE, MARIO}.

The so-called {\it multimedia recommender systems} using such multimodal data of items generate embeddings (\ie, modality features) for each modality of items via pre-trained deep-learning models such as recurrent neural networks~\cite{LSTM} and convolutional neural networks~\cite{ImageNet} beforehand,\footnote{It is not the primary focus of multimedia recommender systems how to generate these modality features via pre-trained deep-learning models.} and then utilize these modality features in addition to user–item interactions. In early studies, most multimedia recommender systems~\cite{VBPR,JRL, VECF, CMF, DeepStyle, CDR, CTR, TopicMF} represent user/item embeddings with textual and visual modality features\footnote{Of course, other modality features may exist; however, current multimedia recommender systems tend to primarily use the textual and visual modality features.} (\ie, multimodal features) and refine these user/item embeddings by learning user--item interactions. This enables multimedia recommender systems to more precisely capture users' preferences for items, thereby improving their recommendation accuracy.

\begin{figure}[t]
\centering
\begin{tikzpicture}
    \small
    \begin{axis}[
        title=Textual Modality,
        title style={at={(0.5,0.94)}},
        ybar=0pt,
        width=4.4cm,
        height=3.0cm,
        bar width=0.35cm,
        bar shift=0pt,
        ylabel={Average},
        ymin=0.08, ymax=0.12,
        symbolic x coords={I--N sim., I--I sim.},
        xtick=data,
        ytick={0.08, 0.09, 0.10, 0.11, 0.12},
        x tick label style={font=\small, /pgf/number format/.cd,fixed,fixed zerofill,precision=2,/tikz/.cd},
        y tick label style={font=\small, /pgf/number format/.cd,fixed,fixed zerofill,precision=2,/tikz/.cd},
        enlarge x limits=0.5,
        ]
        \addplot [ybar, draw=red!60, fill=red!20] coordinates {
            (I--N sim., 0.086630285)
            (I--I sim., 0)
          };
        \addplot [ybar, draw=red!100, fill=red!60] coordinates {
          (I--I sim., 0.10991018)};
    \end{axis}
\end{tikzpicture}
    \hspace{0.3cm}
\begin{tikzpicture}
    \small
    \begin{axis}[
        title=Visual Modality,
        title style={at={(0.5,0.94)}},
        ybar=0pt,
        width=4.4cm,
        height=3.0cm,
        bar width=0.35cm,
        bar shift=0pt,
        ylabel={Average},
        ymin=0.21, ymax=0.33,
        symbolic x coords={I--N sim., I--I sim.},
        xtick=data,
        ytick={0.21, 0.24, 0.27, 0.30, 0.33},
        x tick label style={font=\small, /pgf/number format/.cd,fixed,fixed zerofill,precision=2,/tikz/.cd},
        y tick label style={font=\small, /pgf/number format/.cd,fixed,fixed zerofill,precision=2,/tikz/.cd},
        enlarge x limits=0.5,
        ]
        \addplot [ybar, draw=blue!60, fill=blue!20] coordinates {
        (I--N sim., 0.23256181)
        (I--I sim., 0)
          };
        \addplot [ybar, draw=blue!100, fill=blue!60] coordinates {
          (I--I sim., 0.3009104)};
    \end{axis}
\end{tikzpicture}
\vspace{-0.2cm}
\caption{The average similarities i) between modality features of interacted items and those of non-interacted items (I--N sim.) and ii) between modality features of interacted items (I--I sim.) for textual and visual modalities on the Amazon Women Clothing dataset.}\label{fig:modality}
\vspace{-0.3cm}
\end{figure}
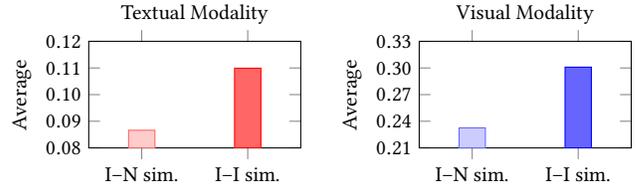
 
A natural question arising is ``why does using multimodal features help capture users' preferences for items?''. Unfortunately, to the best of our knowledge, there has been no study that directly answers this question. Therefore, to clarify this, we focus on users' behavior (\spec, purchase). When a user decides whether to purchase an item, she tends to consider the multimodal data that reveal intrinsic properties of the item. Based on this tendency, we conjecture that there is a difference between the modality features of {\it interacted} items and those of {\it non-interacted} items by each user. To confirm whether or not this conjecture holds in reality, by using the real-world Amazon dataset (\spec, Women Clothing)~\cite{Amazon}, which is widely used in multimedia recommender systems~\cite{VECF, VBPR, MARIO, JRL, LATTICE, lattice_IEEE}, we compute the average cosine similarity between modality features of {\it interacted} items and those of {\it non-interacted} items for each user (the I--N similarity, in short).
For the comparison, we also compute the average cosine similarity between modality features of {\it interacted} items for each user (the I--I similarity, in short). We then take the average of I--N similarities and that of I--I similarities over all users, which is shown in Figure~\ref{fig:modality} w.r.t. each modality.
The results reveal that, although modality features are generated via pre-trained deep-learning models regardless of user--item interactions, the averages of I--I similarities are 26.87\% and 29.39\% higher than those of I--N similarities for textual and visual modalities, respectively.\footnote{This result is statistically significant with $p$-value $\leq 0.001$. We confirmed similar tendencies in other categories of the Amazon dataset as well.}
This result implies that there exists an inherent difference between the multimodal features of interacted items and those of non-interacted items for each user; in other words, it indicates that a user's preference can be revealed by the multimodal features of her interacted items. Thus, we aim to devise a novel approach for judiciously reflecting multimodal features in the final user/item embeddings to more precisely capture users' preferences for items.

\textbf{(Idea 1)} On the other hand, graph convolutional network (GCN) \cite{VGAE,GCN}-based recommender systems have been actively studied for improving the accuracy by {\it explicitly} capturing the {\it collaborative signal}, which is inherent in user--item interactions, to expose {\it behavioral similarity} between users (or items)~\cite{Pinsage,NGCF,LightGCN,LRGCCF,HMLET,siren,CPA-LGC}.
To harness the power of GCN in recommender systems, multimedia recommender systems have also employed GCN~\cite{MMGCN,GRCN,LATTICE}.
Even with the success of GCN-based multimedia recommender systems, we claim that they have the following {\it limitation on capturing users' preferences}: they cannot sufficiently reflect the multimodal features in the final user/item embeddings. This is because the way of {\it neighborhood aggregation} via GCN rather disturbs the effect of reflecting the multimodal features in user/item embeddings. However, it is still effective in improving the accuracy of multimedia recommendation to explicitly capture collaborative signals through GCN. Furthermore, the multimodal features of multi-hop neighbors as well as those of one-hop neighbors will help capture a user's preference. These claims are empirically validated in Section~\ref{sec:preliminaries}. Thus, we would like to design a novel GCN that can incorporate {\it both modality features and collaborative signals} into the final user/item embeddings in a balanced way.

\begin{figure}[t]
\centering
\includegraphics[width=0.38\textwidth]{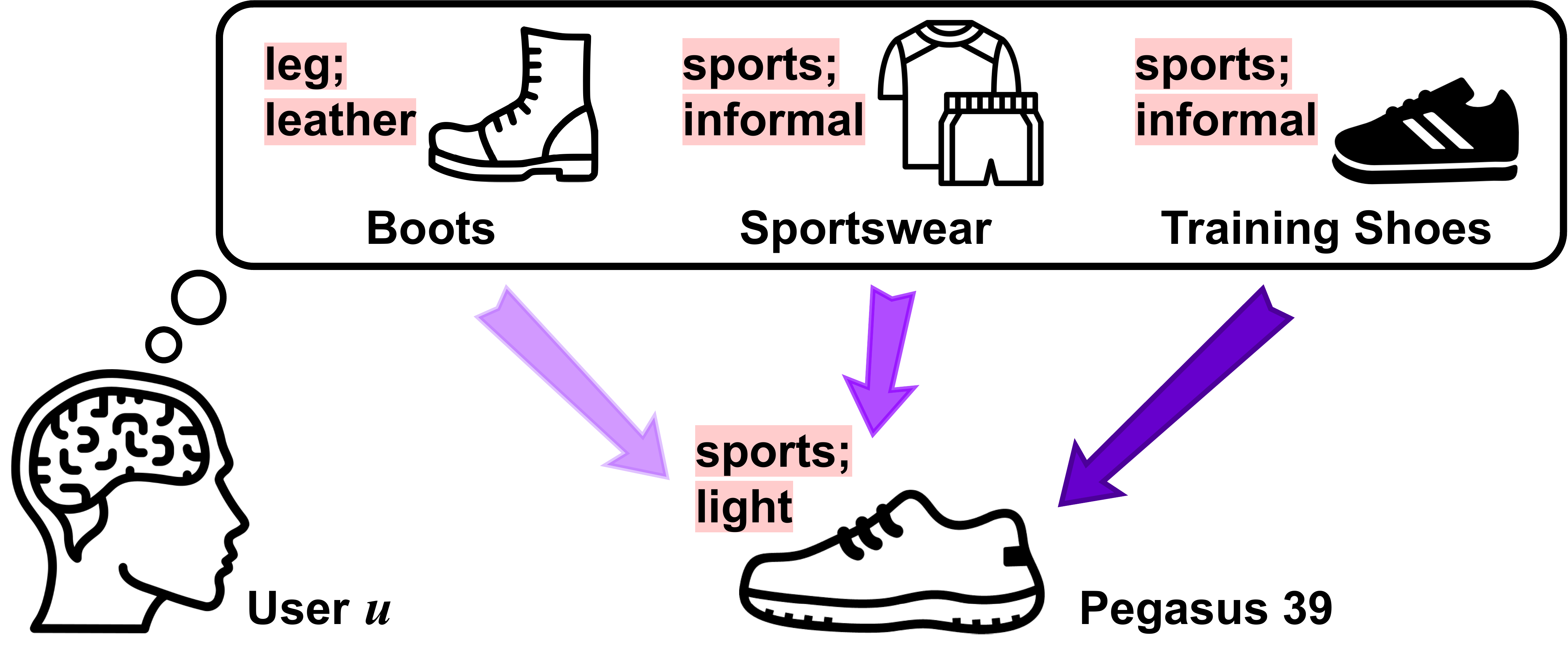}
\vspace{-0.3cm}
\caption{A toy example for user $u$'s interests w.r.t. a target item, Pegasus 39. The darker the color of the arrow, the more user $u$ takes into account her interest in the interacted item when the user $u$ decides whether to purchase Pegasus 39.} \label{fig:example}
\vspace{-0.4cm}
\end{figure}

\textbf{(Idea 2)} There is another {\it challenge on capturing users' preferences for items}. Existing GCN-based multimedia recommender systems represent a user as a {\it single} embedding in which her interests for all interacted items are equally reflected (a {\it user's general embedding}, in short). Then, they utilize the user's general embedding to predict her preference for a non-interacted item (\ie, a target item).
However, when the user decides whether to prefer a target item, the {\it more} the multimodal features of the interacted item are {\it related} to the multimodal features of the target item, the more she will take into account her interest in this (interacted) item.
For example, as illustrated in Figure~\ref{fig:example}, when user $u$, who mainly purchased boots, sportswear, and training shoes, decides whether to purchase the target item (\ie, Pegasus 39), she is likely to take into account her interest in training shoes more than that in sportswear (\textit{resp}. boots). This is because the textual and visual modalities of training shoes are similar to those of Pegasus 39, but the visual modality of sportswear (\textit{resp}. the textual modality of boots) is quite different from that of Pegasus 39.
Therefore, we claim that it is required to generate a {\it target-oriented user embedding} where the interests in interacted items more relevant to the target item w.r.t. multimodal features are more reflected.

In this paper, we aim to capture users' preferences for items more precisely by tackling the two limitations mentioned above. Toward this goal, we propose a novel multimedia recommendation method, named \textbf{\ours}, based on the following two core modules: (M1) {\bf MO}dality-embracing graph convolutional {\bf NE}twork (MeGCN) that can reflect modality features as well as collaborative signals successfully in the final user/item embeddings; (M2) {\bf T}arget-aware attention to generate a user's target-oriented embedding that considers the multimodal features of the target item.

Our contributions are summarized as follows:
\begin{itemize} [leftmargin=*]
\item \textbf{Important Observations}: We point out two limitations of existing GCN-based multimedia recommender systems on precisely capturing users' preferences.
\begin{itemize} [leftmargin=5.0mm]
\item[(L1)] They insufficiently reflect the multimodal features in the final user/item embeddings since they utilize GCN designed to focus only on capturing collaborative signals.
\item[(L2)] They generate a user's general embedding regardless of the multimodal features of the target item and then utilize this (single) user embedding to predict her preference for the target item.
\end{itemize}
\item \textbf{Novel Methodology}: To overcome these two limitations, we propose a novel multimedia recommendation method, named {\ours}, based on MeGCN and target-aware attention.
\item \textbf{Extensive Evaluation}: We validate the rationality and effectiveness of \ours~through extensive experiments on four real-world datasets. Most importantly, {\ours} consistently and dramatically outperforms the state-of-the-art methods, MMGCN~\cite{MMGCN}, GRCN~\cite{GRCN}, LATTICE~\cite{LATTICE}, and MARIO~\cite{MARIO}, by up to 136.31\%, 41.97\%, 31.53\%, and 30.32\%, respectively, in terms of recall@20.
\end{itemize}

\section{Related Work}\label{sec:related_work}
In this section, we review existing multimedia recommender systems and then discuss their limitations.

\vspace{0.1cm}
\noindent\textbf{Multimedia Recommender Systems \underline{Not} Based on GCN}.
Early multimedia recommender systems represent user/item embeddings with multimodal features and refine them by learning user--item interactions~\cite{VBPR, JRL, VECF, CMF, DeepStyle, CDR, CTR, TopicMF}. Specifically, VBPR~\cite{VBPR} represented user/item embeddings with visual modality features and combined them with the matrix factorization (MF) model~\cite{BPR}. JRL~\cite{JRL} represented each user/item as embeddings in three views (\ie, textual modality, visual modality, and numerical rating) and integrated these embeddings into a unified embedding of each user/item through deep-learning techniques.
VECF~\cite{VECF} represented user/item embeddings with visual modality features and additionally utilized textual modality (\spec, user--item reviews) to learn these user/item embeddings.

Recently, beyond representing user/item embeddings with multimodal features, the authors of MARIO~\cite{MARIO} first argued the importance of {\it preserving the modality features} in the final item embeddings in multimedia recommendation. To this end, they proposed a novel modality preservation (MP) loss.
To demonstrate the effectiveness of the MP loss, they defined the notion of {\it average difference} w.r.t. modality $m$ ($\text{avg.diff}_m$, in short) to quantify how much each item loses its modality feature in its final embedding.
Formally, $\text{avg.diff}_m$ can be expressed as follows:
$\forall m \in \{t,v\}$,
\small
\begin{equation}
    \text{avg.diff}_m 
    = \frac{1}{\vert \mathcal{I} \vert(\vert \mathcal{I} \vert -1)}\sum_{\substack{i,j\in\mathcal{I},\\ i\neq j}}\bigg|\frac{\tilde{\textbf{{e}}}_{i,m}^\intercal \cdot \tilde{\textbf{e}}_{j,m}}{\Vert \tilde{\textbf{e}}_{i,m} \Vert \Vert \tilde{\textbf{e}}_{j,m} \Vert} - \frac{\textbf{e}_{i,m}^\intercal \cdot \textbf{e}_{j,m}}{\Vert \textbf{e}_{i,m} \Vert \Vert \textbf{e}_{j,m} \Vert}\bigg|,
\label{eq:1}
\end{equation}
\normalsize
where $t$ and $v$ denote textual and visual modalities, respectively; $\mathcal{I}$ denotes the set of all items; $\tilde{\textbf{e}}_{i,m} \in \mathbb{R}^{d_m}$ (\textit{resp}. $\textbf{e}_{i,m} \in \mathbb{R}^{d}$) denotes the modality feature (\textit{resp}. the final item embedding) of item $i$ w.r.t. modality $m$; $d_m$ denotes the dimensionality of modality feature w.r.t. modality $m$; $d$ denotes the dimensionality of the final item embedding. The smaller the score of this measure, the better the modality features are preserved in the final item embeddings.

\begin{table}[t]
\centering
\caption{GCN-based methods w.r.t. GCN techniques}
\vspace{-0.3cm}
\label{table:relatedwork}
\small
\renewcommand{\arraystretch}{1.15}
\resizebox{.48\textwidth}{!}{
\begin{tabular}{c|c|c|cccc}
\toprule
\textbf{} & \textbf{GCN} & \textbf{LightGCN} & \textbf{MMGCN} & \textbf{GRCN} & \textbf{LATTICE} & \textbf{\ours} \\ \midrule
\textbf{Non-linear Prop.} & O & X & O  & X  & X & X \\
\textbf{Linear Prop.} & X & O & X  & O  & O & O \\
\textbf{Self-Conn.} & O & X & O  & X  & X & O \\
\textbf{Layer Comb.} & X & O & X  & O  & O & X \\
\bottomrule
\end{tabular}
}
\vspace{-0.5cm}
\end{table}

\vspace{0.1cm}
\noindent\textbf{Multimedia Recommender Systems Based on GCN}.
While taking advantage of explicitly capturing collaborative signals in terms of improving the accuracy, GCN-based multimedia recommender systems mainly utilize multimodal features for one of the following two purposes: (P1) representing the user/item embeddings; (P2) refining the relationships between nodes.
For (P1), MMGCN~\cite{MMGCN} first built a user--item interaction bipartite graph and {\it initialized item embeddings} based on their modality features and user embeddings randomly w.r.t. each modality. Then, it employed GCN based on not only {\it non-linear propagation}, which uses a weight matrix and an activation function~\cite{HMLET}, but also {\it self-connection}, which adds the node's own embedding on {\it each} layer, independently w.r.t. each modality.
For (P2), GRCN~\cite{GRCN} and LATTICE~\cite{LATTICE} refined user--item relationships and item--item relationships, respectively. Specifically, GRCN utilized multimodal features to refine the noisy edges of the user--item interaction bipartite graph. Thereafter, it randomly initialized the user/item embeddings of the refined graph, and then employed GCN based on not only {\it linear propagation}, which does not use a weight matrix and an activation function~\cite{HMLET}, but also {\it layer combination} that aggregates embeddings in {\it all} layers of each node at once.
LATTICE constructed a latent item--item graph using multimodal features and employed GCN based on {\it linear propagation} to learn the item embeddings randomly initialized in the constructed graph. Then, it combined these item embeddings with collaborative filtering (CF) methods (\spec, LightGCN~\cite{LightGCN} that employs GCN based on not only {\it linear propagation} but also {\it layer combination}). We summarize the above GCN-based methods, along with the original GCN~\cite{GCN} and LightGCN for comparison, in Table~\ref{table:relatedwork} w.r.t. the mentioned GCN techniques.

\vspace{0.1cm}
\noindent\textbf{Discussions}.
The aforementioned methods represented each user as a {\it single} embedding regardless of the multimodal features of the target item. Thus, they cannot accommodate the fact that, when a user decides whether to prefer a given target item, the more the multimodal features of the interacted item by the user are related to the multimodal features of the target item, the more she is to take into account the interest in this (interacted) item.
Furthermore, except for MARIO, existing methods were not aware of the importance of preserving modality features in the final item embeddings. On the other hand, MARIO recently showed that preserving modality features in the final item embeddings helps improve the accuracy of multimedia recommendation, thereby leading to the state-of-the-art performance in multimedia recommendation.
However, it cannot capture users’ preferences that can be revealed by the multimodal features of {\it multi-hop neighbors}, since it is not based on GCN but based on simple MF.
Thus, as long as GCN is utilized to successfully reflect multimodal features in the final user/item embeddings, there is still a room for improvement in the accuracy of multimedia recommendation. However, it is challenging to sufficiently reflect multimodal features into the final user/item embeddings by using {\it existing} GCNs due to its way of neighborhood aggregation.
\section{Preliminaries}\label{sec:preliminaries}
In this section, we validate the following two claims: (C1) existing GCN-based multimedia recommender system (\spec, MMGCN \cite{MMGCN}) that utilizes multimodal features for representing the user/item embeddings insufficiently reflects multimodal features into the final user/item embeddings;\footnote{As mentioned earlier, GRCN~\cite{GRCN} and LATTICE~\cite{LATTICE} utilize multimodal features not for representing the user/item embeddings, but for refining the relationships between nodes. Thus, we exclude them from this claim.} (C2) nevertheless, GCN is still effective in improving the accuracy of multimedia recommendation. To this end, we conducted experiments by using variants of MMGCN equipped with different combinations of the {\it GCN techniques} (see Table~\ref{table:relatedwork}) on the real-world Amazon Women Clothing dataset.\footnote{The results on other categories of the Amazon dataset showed similar tendencies to those on Women Clothing.}
For a fair comparison, we tuned the hyperparameters of MMGCN via extensive grid search and used the tuned best hyperparameters in all variants. Also, we report the average of values obtained by the five independent evaluations for each measure.

\vspace{1mm}
\noindent\textbf{Experimental Analysis on Design of GCN}.
To validate our claim (C1), we compared MMGCN, which employs GCN based on {\it non-linear propagation} and {\it self-connection}, with its following three variants depending on the GCN techniques: (V1) MMGCN-n: a variant of MMGCN employing linear propagation instead of non-linear propagation to investigate the effects of non-linear and linear propagations; (V2) MMGCN-s: a variant of MMGCN removing self-connection to examine the effect of self-connection; (V3) MMGCN-s+l: a variant of MMGCN employing a layer combination instead of self-connection to examine the effect of layer combination.\footnote{The layer combination is known as a similar technique to self-connection~\cite{LightGCN}.}
For empirical validation, we measured both the accuracy (\spec, recall@20) and $\text{avg.diff}_m$ (defined in Eq.~\eqref{eq:1}) for $m\in \{t,v\}$ to quantify how well the modality features are reflected in the final item embeddings\footnote{This measure is defined only for items rather than users because users do not have multimodal features.} generated by each method, same as in MARIO~\cite{MARIO}.
Recall that the smaller the score of $\text{avg.diff}_m$, the better the modality $m$'s features are reflected in the final item embeddings.

\begin{table}[t]
\centering
\caption{Experimental results of MMGCN and its three variants on the Amazon Women Clothing dataset. The best and second-best results in each column (\ie, measure) are highlighted by bold and underline, respectively. All differences are statistically significant with $p$-value $\leq 0.001$.}
\vspace{-0.2cm}
\small
\resizebox{.36\textwidth}{!}{
\begin{tabular}{c|cc|c}
    \toprule 
    \multirow{1}{*} & \multicolumn{2}{c|}{\textbf{Average differences}} & \textbf{Accuracy} \\ \cline{2-4} 
    \multirow{1}{*}{\textbf{Methods}} & \textbf{avg.diff$_t$} & \textbf{avg.diff$_v$} & \textbf{recall@20} \\ \hline 
    \multicolumn{1}{c|}{\textbf{MMGCN}}     & \underline{0.7559} & \underline{0.5992} & \underline{0.0419} \\ \hline
    \multicolumn{1}{c|}{\textbf{MMGCN-n}}     & \textbf{0.1547} & \textbf{0.1820}   & \textbf{0.0597} \\
    \multicolumn{1}{c|}{\textbf{MMGCN-s}}     & 0.8309 & 0.6954   & 0.0123 \\
    \multicolumn{1}{c|}{\textbf{MMGCN-s+l}}     & 0.8010 & 0.6870  & 0.0341 \\ \bottomrule
\end{tabular}
}
\label{table:pre}
\end{table}

\definecolor{purple}{rgb}{0.56, 0.0, 1.0}

\begin{figure}[t]
\centering
\begin{tikzpicture}
    \footnotesize
    \begin{axis}[
        xbar,
        width=6cm,
        height=2.8cm,
        bar width=0.3cm,
        bar shift=0pt,
        xlabel={Recall@20},
        xmin=0, xmax=0.05,
        symbolic y coords={MMGCN-GCN, MMGCN},
        ytick=data,
        xtick={0, 0.01, 0.02, 0.03, 0.04, 0.05},
        x tick label style={font=\footnotesize, /pgf/number format/.cd,
                scaled x ticks = false,
                fixed,
                precision=2
        },
        enlarge y limits=0.5,
        ]
        \addplot [draw=purple!100, fill=purple!70] coordinates {
            (0.002,MMGCN-GCN)
            (0.000,MMGCN)
          };
        \addplot [draw=purple!70, fill=purple!25] coordinates {
          (0.0419,MMGCN)};
    \end{axis}
\end{tikzpicture}
\vspace{-0.4cm}
\caption{Accuracies of MMGCN and MMGCN-GCN on the Amazon Women Clothing dataset. The improvement is statistically significant with $p$-value $\leq 0.001$.}\label{fig:gcn}
\vspace{-0.4cm}
\end{figure}
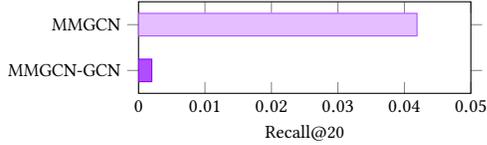

Table~\ref{table:pre} summarizes $\text{avg.diff}_m$ for $m\in \{t,v\}$ and the recall@20 for MMGCN and its three variants, respectively. From Table~\ref{table:pre}, our empirical findings are summarized as follows:
\begin{longenum}
\item \textbf{MMGCN vs. MMGCN-n}: The use of a weight matrix and an activation function in the propagation step (\ie, non-linear propagation) extremely disturbs reflecting the modality features into the final embeddings due to the {\it distortion} of the modality features in its nature;
\item \textbf{MMGCN vs. MMGCN-s}: The use of self-connection is effective in reflecting modality features into the final embeddings since it represents an item's embedding {\it including its own modality feature};
\item \textbf{MMGCN vs. MMGCN-s+l}: The use of layer combination instead of self-connection weakens reflecting the modality features into the final embeddings. This is because self-connection is analytically known to reflect the initial embedding (\ie, modality feature) more than layer combination~\cite{LightGCN};
\item \textbf{Relationship between avg.diff$_m$ and accuracy}: In all cases, the lower the score of avg.diff$_m$, the higher the accuracy. In other words, reflecting modality features into the final embeddings to a {\it sufficient extent} leads to improving the accuracy.
\end{longenum}

From the first empirical finding, we validated our claim (C1). Furthermore, we demonstrated that, in the propagation step, it is important {\it not to distort modality features} but to reflect them sufficiently into the final embeddings for improving the accuracy.

\vspace{1mm}
\noindent\textbf{Effect of GCN}.
If {\it only} the more reflection of modality features in the final embeddings leads to the higher accuracy in multimedia recommendation, someone could consider not to use GCN in the design of her multimedia recommender system, since the neighborhood aggregation in GCN is not indeed beneficial. However, we claim that the neighborhood aggregation helps explicitly capture collaborative signals and the multimodal features of multi-hop neighbors as well as those of one-hop neighbors help capture a user’s preference, thus enabling to improve the accuracy (\ie, our claim (C2)).
To validate this claim, we compared MMGCN with its variant, namely MMGCN-GCN, which is MMGCN without GCN (\ie, the case where the number of GCN-layers is zero).

Figure~\ref{fig:gcn} shows the accuracies of MMGCN and MMGCN-GCN in terms of recall@20.
MMGCN-GCN shows a much lower accuracy than that of MMGCN. This result implies that GCN plays a crucial role in obtaining a high accuracy of multimedia recommendation.

\vspace{1mm}
Inspired by these insights, we propose a novel GCN method {\it suitable for multimedia recommendation} based on {\it linear propagation} and {\it self-connection}, as elaborated in detail in Section~\ref{sec:components}.
\section{\textsc{\textsf{{MONET}}}: Proposed Method}\label{sec:approach}

\begin{figure*}[t!]
\centering
\includegraphics[width=1.0\textwidth]{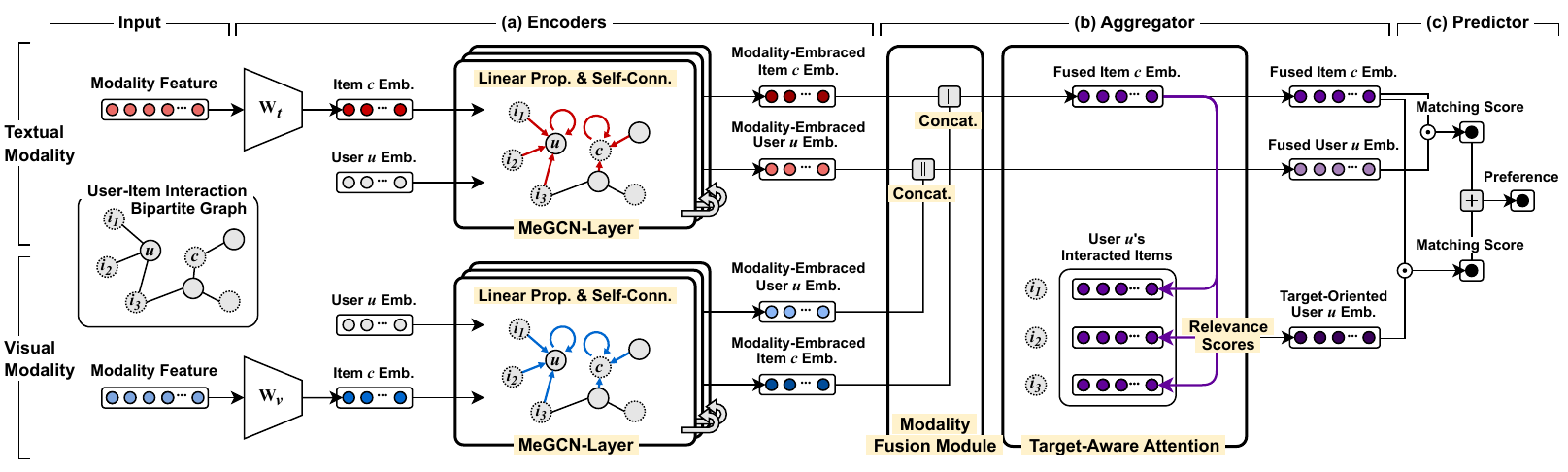}
\vspace{-0.6cm}
\caption{Overview of {\ours} composed of three key components: (a) encoders based on MeGCN; (b) an aggregator based on target-aware attention; (c) a predictor utilizing both general embedding and target-oriented embedding of a user.} \label{fig:monet}
\vspace{-0.3cm}
\end{figure*}

In this section, we elaborate on our proposed method, named {\ours}, based on the MeGCN and target-aware attention.

\subsection{Problem Definition}\label{sec:problem_definition}
Let $u\in\mathcal{U}$ and $i\in\mathcal{I}$ denote a user and an item, respectively, where $\mathcal{U}$ and $\mathcal{I}$ denote the sets of all users and all items, respectively; $\mathcal{N}_u\subset\mathcal{I}$ denotes a set of items interacted by user $u$. Here, item $i$ can have multiple modalities (\eg, textual, visual, and acoustic modalities). In this paper, we assume that each item has textual and visual modalities as in~\cite{VECF, LATTICE, MARIO}.\footnote{
If items’ additional modalities exist, they can be easily incorporated into {\ours}.} Thus, each item has (pre-trained) modality feature $\tilde{\textbf{e}}_{i,m}\in\mathbb{R}^{d_m}$ w.r.t. modality $m\in\{t,v\}$, where $t$ and $v$ denote textual and visual modalities, respectively; $d_m$ indicates the dimensionality of modality features w.r.t. modality $m$.
The goal of multimedia recommendation is to identify the top-$N$ items that user $u$ is most likely to prefer among her non-interacted items $c\in\mathcal{I}{\setminus}\mathcal{N}_u$ by using not only the user--item interactions but also the multimodal features of items. Table~\ref{table:notations} summarizes the key notations used in this paper.

\subsection{Key Components of \ours}\label{sec:components}
Now, we describe three key components (\ie, encoders, an aggregator, and a predictor) of \ours~in detail. The schematic overview of \ours~is illustrated in Figure~\ref{fig:monet}.

\vspace{1mm}
\noindent\textbf{Encoders}.
\ours~first represents user--item interactions as a user--item interaction bipartite graph $\mathcal{G}=\left(\mathcal{V},\mathcal{E}\right)$, where $\mathcal{V}=\left(\mathcal{U} \cup \mathcal{I}\right)$ and $\mathcal{E}=\{(u,i) \mid u \in \mathcal{U},i \in \mathcal{N}_u\}$ denote the sets of nodes and edges, respectively. Given a user--item interaction bipartite graph $\mathcal{G}$, \ours~randomly initializes user embedding $\textbf{e}_{u,m}^0\in\mathbb{R}^d$ differently for each modality $m$, where $d$ is the dimensionality of the embedding. For item $i$, \ours~transforms modality $m$'s feature $\tilde{\textbf{e}}_{i,m}\in\mathbb{R}^{d_m}$ into initial item embedding $\textbf{e}_{i,m}^0\in\mathbb{R}^d$ that represents its high-level feature~\cite{GRCN, LATTICE}, which is expressed as follows:
$\forall m \in \{t, v\}$,
\vspace{-0.1cm}
\small
\begin{equation}
    \textbf{e}_{i,m}^0=\textbf{W}_m\tilde{\textbf{e}}_{i,m}+\textbf{b}_m,
\label{eq:2}
\end{equation}
\normalsize
where $\textbf{W}_m\in\mathbb{R}^{d{\times}d_m}$ and $\textbf{b}_m\in\mathbb{R}^d$ denote a trainable weight matrix and a bias vector, respectively.

Recall that it is important to {\it sufficiently reflect modality features as well as collaborative signals into user/item embeddings} for obtaining high accuracy.
To achieve this, we design MeGCN based on {\it linear propagation} and {\it self-connection}, which is inspired from our empirical findings in Section~\ref{sec:preliminaries}. \ours~applies this MeGCN to learn user/item embeddings independently for each modality $m$.
Formally, in the $l$-th layer, item embedding $\textbf{e}_{i,m}^l$ for modality $m$ via MeGCN can be expressed as follows:
$\forall m \in \{t, v\}$,
\vspace{-0.1cm}
\small
\begin{equation}
    \textbf{e}_{i,m}^l= \sum_{u \in \mathcal{N}_i}\frac{1}{\sqrt{\vert \mathcal{N}_u \vert \vert \mathcal{N}_i \vert}}\textbf{e}_{u,m}^{l-1} + \alpha \times \textbf{e}_{i,m}^{l-1},
\label{eq:3}
\end{equation}
\normalsize
where $\mathcal{N}_i\subset\mathcal{U}$ denotes a set of users who interacted with item $i$ and $\alpha$ denotes the coefficient of self-connection that controls the degree of reflection of a modality feature. The user embedding $\textbf{e}_{u,m}^l$ at the $l$-th layer can also be represented similarly as in Eq.~\eqref{eq:3}. Here, we highlight that MeGCN does not normalize the second term (corresponding to self-connection) in Eq.~\eqref{eq:3}, unlike the first term (corresponding to collaborative signals), in order to basically reflect the embeddings of neighbors and its own embedding to similar degrees in the $l$-th embedding.\footnote{We experimented with a variant of MeGCN that normalizes the self-connection term along with collaborative signals term, but the current version of MeGCN, which does not normalize the self-connection term, showed higher accuracy.} We will provide the sensitivity analysis for $\alpha$ in Section~\ref{sec:evaluation} to analyze the relationship between collaborative signals and modality features.

After stacking $L$ MeGCN-layers, \ours~obtains the $L$-th user (\textit{resp}. item) embedding $\textbf{e}_{u,m}^L$ (\textit{resp}. $\textbf{e}_{i,m}^L$) for modality $m$, and uses it as the resulting modality-embraced user (\textit{resp}. item) embedding $\textbf{e}_{u,m}$ (\textit{resp}. $\textbf{e}_{i,m}$) w.r.t. modality $m$.
By utilizing the modality-embraced embeddings $\textbf{e}_{u,m}$ and $\textbf{e}_{i,m}$, \ours~can not only well represent item $i$ with its modality $m$'s feature but also more precisely capture user $u$'s preferences on items since her preferences can be revealed by the modality features as shown in Figure~\ref{fig:modality}.

\begin{table}[t]
\normalsize
  \centering
  \caption{Key notations used in this paper}  
  \vspace{-0.3cm}
\small
\resizebox{.48\textwidth}{!}{
    \begin{tabular}{p{0.15\linewidth}|p{0.85\linewidth}} \toprule
    \hfil \textbf{Notation}                                                           & \hfil \textbf{Description} \\ \midrule
    \hfil $t, v$                                                                         &  Textual and visual modalities \\  \midrule
    \hfil $\tilde{\textbf{e}}_{i,m}$ & Modality $m$'s feature of item $i$ \\
    \hfil $\textbf{e}_{u,m}^l, \textbf{e}_{i,m}^l$              &  Embeddings of user $u$ and item $i$ w.r.t. modality $m$ at the $l$-th layer \\
    \hfil $\textbf{e}_{u,m}, \textbf{e}_{i,m}$ & Modality-embraced embeddings of user $u$ and item $i$ w.r.t. modality $m$ \\
    \hfil $\textbf{e}_{u}, \textbf{e}_{i}$              & Fused embeddings of user $u$ and item $i$ \\
    \hfil $\textbf{e}_{u}^{c}$ & Target-oriented embedding of user $u$ for target item $c$ \\ \midrule
    \hfil $d$              &  Dimensionality of the embedding \\
    \hfil $d_m$              & Dimensionality of feature w.r.t. modality $m$ \\ \midrule
    \hfil $a_u^{c,i}$ & Relevance score between target item $c$ and user $u$'s interacted item $i$ \\
    \hfil $\hat{y}_{u,c}^g$ & Matching score between fused embeddings of user $u$ and target item $c$ \\
    \hfil $\hat{y}_{u,c}^o$ & Matching score between target-oriented embedding of user $u$ and fused embedding of target item $c$ \\
    \hfil $\hat{y}_{u,c}$ & Preference of user $u$ for target item $c$\\ \midrule
    \hfil $\alpha$ & Coefficient of self-connection that controls the
degree of reflection of a modality feature \\ 
    \hfil $\beta$ &  Parameter balancing between two matching scores $\hat{y}_{u,c}^g$ and $\hat{y}_{u,c}^o$  \\
    \bottomrule
  \end{tabular}
  }
  \vspace{-0.4cm}
  \label{table:notations}
\end{table}

\vspace{1mm}
\noindent\textbf{Aggregator}.
Then, \ours~ generates a fused user (\textit{resp}. item) embedding $\textbf{e}_u\in\mathbb{R}^{2d}$ (\textit{resp}. $\textbf{e}_i\in\mathbb{R}^{2d}$) via a modality fusion module that {\it concatenates} the modality-embraced user (\textit{resp}. item) embeddings $\textbf{e}_{u,m}$ (\textit{resp}. $\textbf{e}_{i,m}$) over $\forall m \in \{t, v\}$ (\ie, $\textbf{e}_u=[ \textbf{e}_{u,t}||\textbf{e}_{u,v}] \text{ and } \textbf{e}_i=[\textbf{e}_{i,t}||\textbf{e}_{i,v}]$).
This fusion strategy is simple but effective in more precisely capturing user $u$'s preferences on items since it generates a fused user (\textit{resp}. item) embedding $\textbf{e}_u\in\mathbb{R}^{2d}$ (\textit{resp}. $\textbf{e}_i\in\mathbb{R}^{2d}$) by not distorting, but preserving the enriched information of modality-embraced embeddings $\textbf{e}_{u,m}$ (\textit{resp}. $\textbf{e}_{i,m}$) for $\forall m \in \{t, v\}$.\footnote{The effectiveness of this way of fusion is empirically validated in \url{https://sites.google.com/hanyang.ac.kr/monet-wsdm2024}.}
In this process, we also considered employing {\it additional user/item CF embeddings} obtained by learning only user--item interactions for achieving higher accuracy. However, we did not take this design choice finally because our experiment for the validation of this rather showed its negative effect on accuracy improvement.
We conjecture that this result was obtained because modality-embraced embeddings $\textbf{e}_{u,m}$ and $\textbf{e}_{i,m}$ for $\forall m \in \{t, v\}$ already capture the user--item interactions (\ie, users' behaviors in CF) well via MeGCN.

Subsequently, we note that fused user embedding $\textbf{e}_u$ was generated regardless of the multimodal features of target item $c\in\mathcal{I}{\setminus}\mathcal{N}_u$. That is, fused user embedding $\textbf{e}_u$ corresponds to user $u$'s {\it general} embedding in which the interests in her all interacted items ($\forall i \in \mathcal{N}_u$) are {\it equally} reflected.
However, as illustrated in Figure~\ref{fig:example}, in order to more precisely capture user $u$'s preference for target item $c$, it is required to generate user $u$'s {\it target-oriented} embedding in such a way that the interests in her interacted items $i \in \mathcal{N}_u$ more relevant to the target item $c$ w.r.t. multimodal features are more reflected.
Towards this end, we propose target-aware attention to generate target-oriented user embedding $\textbf{e}_u^c\in\mathbb{R}^{2d}$ by aggregating user $u$’s interests in interacted items $i \in \mathcal{N}_u$ based on the \textit{relevance score} between multimodal features of interacted item $i$ and the target item $c$.\footnote{All of a user's non-interacted items are regarded as her target items. Nevertheless, {\ours} is applicable to the industry-scale real-world recommender systems because most of them reduce the number of target items to around hundreds by performing a candidate generation step~\cite{Youtube, Two-tower, AiRS}. It is not covered in this paper, since this falls outside the scope of our research.} Specifically, \ours~first calculates the {\it relevance score} $a_u^{c,i}$ between the two fused embeddings, $\textbf{e}_c$ of target item $c$ and $\textbf{e}_i$ of user $u$'s interacted item $i$, as follows:
$\forall i \in \mathcal{N}_u$,
\small
\begin{equation}
    a_u^{c,i}=\frac{\exp(\textbf{e}_c^\intercal \cdot \textbf{e}_i)}{\sum_{j{\in}\mathcal{N}_u}{\exp(\textbf{e}_c^\intercal \cdot \textbf{e}_j)}}.
\label{eq:5}
\end{equation}
\normalsize
Then, \ours~aggregates the fused item embeddings $\textbf{e}_i$ based on the relevance scores $a_u^{c,i}$ for $\forall i \in \mathcal{N}_u$ to obtain the target-oriented user embedding $\textbf{e}_u^c$, as follows:

\small
\begin{equation}
    \textbf{e}_u^c=\sum_{i{\in}\mathcal{N}_u}{a_u^{c,i}\times\textbf{e}_i}.
\label{eq:6}
\end{equation}
\normalsize

\vspace{1mm}
\noindent\textbf{Predictor}.
Lastly, \ours~predicts user $u$'s preference for target item $c$.
It is worth noting that fused user embedding $\textbf{e}_u$ (\ie, her {\it general} embedding) is still useful for representing the user $u$ since it contains user $u$'s {\it general interest}.
Thus, {\ours} leverages {\it both} user $u$'s fused embedding $\textbf{e}_u$ and target-oriented embedding $\textbf{e}_u^c$.

Specifically, \ours~first computes the matching score $\hat{y}_{u,c}^g$  between the two fused embeddings $\textbf{e}_u$ of user $u$ and $\textbf{e}_c$ of target item $c$ via the dot-product (\ie, $\hat{y}_{u,c}^g=\textbf{e}_u^\intercal \cdot \textbf{e}_c$).
Next, \ours~computes the matching score $\hat{y}_{u,c}^o$ between target-oriented embedding $\textbf{e}_u^c$ of user $u$ and fused embedding $\textbf{e}_c$ of target item $c$ via the dot-product (\ie, $\hat{y}_{u,c}^o={\textbf{e}_u^c}^\intercal \cdot \textbf{e}_c$).
Lastly, \ours~predicts user $u$'s preference $\hat{y}_{u,c}$ for target item $c$ by leveraging both matching scores $\hat{y}_{u,c}^g$ and $\hat{y}_{u,c}^o$, as follows:
\vspace{-0.1cm}
\small
\begin{equation}
    \hat{y}_{u,c}=(1-\beta)\times\hat{y}_{u,c}^g+\beta\times\hat{y}_{u,c}^o,
\label{eq:9}
\end{equation}
\normalsize
where $\beta\text{ }{\in}\text{ }(0,1)$ denotes a parameter balancing between two matching scores $\hat{y}_{u,c}^g$ and $\hat{y}_{u,c}^o$. By leveraging both matching scores $\hat{y}_{u,c}^g$ and $\hat{y}_{u,c}^o$, {\ours} can more precisely capture user $u$'s preference for target item $c$ since the two matching scores $\hat{y}_{u,c}^g$ and $\hat{y}_{u,c}^o$ capture different but important aspects of user $u$'s preference, which will be empirically validated in Section~\ref{sec:evaluation}.

\subsection{Optimization}\label{sec:optimization}
We adopt the Bayesian Personalized Ranking (BPR) loss~\cite{BPR}, which is widely used for optimizing multimedia recommender systems~\cite{VBPR, MMGCN, GRCN, LATTICE, MARIO}, to learn the trainable parameters of \ours, so that the preference for the interacted item of a user is likely to be higher than that for the non-interacted item of a user, as follows:
\vspace{-0.1cm}
\small
\begin{equation}
    \mathcal{L}=-\sum_{u\in\mathcal{U}}\sum_{i{\in}\mathcal{N}_u}\sum_{j{\in}\mathcal{I}{\setminus}\mathcal{N}_u}{\ln\sigma(\hat{y}_{u,i}-\hat{y}_{u,j})}+\lambda\Vert\boldsymbol{\uptheta}\Vert_2^2,
\label{eq:10}
\end{equation}
\normalsize
where $\sigma(\cdot)$ denotes a sigmoid function, $\lambda$ denotes a regularization weight, and $\boldsymbol{\uptheta}$ denotes trainable parameters of \ours.

\section{Evaluation}\label{sec:evaluation}
In this section, we conduct extensive experiments to answer the following key research questions (RQs):
\begin{itemize}[leftmargin=*]
\item \textbf{RQ1}: Does \ours~provide a more-accurate top-$N$ recommendation than state-of-the-art GCN-based recommender systems and state-of-the-art multimedia recommender systems?
\item \textbf{RQ2}: Does \ours~reflect modality features into the final embeddings better than other methods?
\item \textbf{RQ3}: Are \ours's two core modules (MeGCN and target-aware attention) effective in improving the accuracy?
\item \textbf{RQ4}: How sensitive is the accuracy of \ours~to hyperparameters $\alpha$ and $\beta$?
\end{itemize}

\vspace{-0.3cm}
\subsection{Experimental Settings}\label{sec:setting}
\noindent\textbf{Datasets}. We used the following four categories of the real-world Amazon dataset \cite{Amazon}, which is publicly available and widely used in the studies on multimedia recommendation~\cite{VBPR, JRL, VECF, LATTICE, MARIO, lattice_IEEE}: Women Clothing, Men Clothing, Beauty, and Toys \& Games. They all contain both textual and visual modalities as well as user--item interactions, where each user/item has at least five interactions. Table~\ref{table:datasets} provides some statistics of the four datasets.

\begin{table}[t]
\raggedright
\caption{Dataset statistics}
\vspace{-0.3cm}
\label{table:datasets}
\small
\resizebox{.48\textwidth}{!}{
\begin{tabular}{crrrr}
\toprule
\textbf{Dataset} & \textbf{\# Users} & \textbf{\# Items} & \textbf{\# Interactions} & \textbf{Sparsity} \\ \midrule
\textbf{Women Clothing} & 19,244  & 14,596  & 135,326  & 99.95\%  \\
\textbf{Men Clothing} & 4,955  & 5,028  & 32,363  & 99.87\%  \\ 
\textbf{Beauty} & 22,363  & 12,101  & 198,502  & 99.93\%  \\
\textbf{Toys \& Games} & 19,412  & 11,924  & 167,597  & 99.93\%  \\
\bottomrule
\end{tabular}
}
\vspace{-0.4cm}
\end{table}
\begin{table*}[t!]
\centering
\caption{Accuracies of seven competitors and \ours~on four datasets. The improvements of \ours~over the best competitors are all statistically significant with $p$-value $\leq 0.001$.}\label{table:sota}
\vspace{-0.2cm}
\renewcommand{\arraystretch}{1.2}
\resizebox{1.0\textwidth}{!}{
\begin{tabular}{ccccccccccccc}
\toprule
\textbf{Datasets} & \multicolumn{3}{c}{\textbf{Women Clothing}}                                   & \multicolumn{3}{c}{\textbf{Men Clothing}} & \multicolumn{3}{c}{\textbf{Beauty}} & \multicolumn{3}{c}{\textbf{Toys \& Games}}  \\ \cmidrule(lr){1-1} \cmidrule(lr){2-4} \cmidrule(lr){5-7} \cmidrule(lr){8-10} \cmidrule(lr){11-13}
\textbf{Measures}                       & \textbf{P@20}            & \textbf{R@20}            & \textbf{NDCG@20}     & \textbf{P@20}            & \textbf{R@20}            & \textbf{NDCG@20}      & \textbf{P@20}            & \textbf{R@20}            & \textbf{NDCG@20}        & \textbf{P@20}            & \textbf{R@20}            & \textbf{NDCG@20} \\ \midrule
\multicolumn{1}{c|}{\textbf{NGCF} (SIGIR'19)}                   & 0.0017         & 0.0344          & \multicolumn{1}{c|}{0.0151}       & 0.0018         & 0.0366          & \multicolumn{1}{c|}{0.0146}       & 0.0041         & 0.0724          & \multicolumn{1}{c|}{0.0328}        & 0.0030         & 0.0567          & 0.0254 \\
\multicolumn{1}{c|}{\textbf{LightGCN} (SIGIR'20)}                   & 0.0032         & 0.0635          & \multicolumn{1}{c|}{0.0301}       & 0.0021         & 0.0412          & \multicolumn{1}{c|}{0.0184}       & 0.0066         & 0.1181          & \multicolumn{1}{c|}{0.0565}        & 0.0059         & 0.1116          & 0.0551   \\ \hline
\multicolumn{1}{c|}{\textbf{VBPR} (AAAI'16)}               &  0.0025        & 0.0483          & \multicolumn{1}{c|}{0.0213}       & 0.0027         & 0.0532          & \multicolumn{1}{c|}{0.0220}       &    0.0046      & 0.0808          & \multicolumn{1}{c|}{0.0395}        & 0.0040         & 0.0745          & 0.0355       \\
\multicolumn{1}{c|}{\textbf{MMGCN} (MM'19)}                   & 0.0021         & 0.0419          & \multicolumn{1}{c|}{0.0180}       & 0.0020         & 0.0389          & \multicolumn{1}{c|}{0.0152}       & 0.0047         & 0.0812          & \multicolumn{1}{c|}{0.0367}        & 0.0033         & 0.0619          & 0.0277   \\
\multicolumn{1}{c|}{\textbf{GRCN} (MM'20)}                   &  0.0035        & 0.0697           & \multicolumn{1}{c|}{0.0302}       & 0.0028         & 0.0562          & \multicolumn{1}{c|}{0.0234}       & 0.0072         & 0.1268          & \multicolumn{1}{c|}{0.0615}        & 0.0065         & 0.1210           &  0.0585 \\
\multicolumn{1}{c|}{\textbf{LATTICE} (MM'21)}                   & 0.0038         & 0.0753          & \multicolumn{1}{c|}{0.0348}       & 0.0031         & 0.0612          & \multicolumn{1}{c|}{0.0273}       & \underline{0.0076}         & \underline{0.1362}          & \multicolumn{1}{c|}{\underline{0.0657}}        & \underline{0.0068}         & \underline{0.1284}          & \underline{0.0629}   \\
\multicolumn{1}{c|}{\textbf{MARIO} (CIKM'22)}                   & \underline{0.0039}         & \underline{0.0760}          & \multicolumn{1}{c|}{\underline{0.0349}}       & \underline{0.0035}         & \underline{0.0688}          & \multicolumn{1}{c|}{\underline{0.0304}}       & 0.0071         & 0.1275          & \multicolumn{1}{c|}{0.0612}        & 0.0064         & 0.1212          & 0.0587   \\ \hline
\multicolumn{1}{c|}{\textbf{\ours}}                  & \textbf{0.0050}         & \textbf{0.0990}          & \multicolumn{1}{c|}{\textbf{0.0450}}       & \textbf{0.0045}         & \textbf{0.0895}          & \multicolumn{1}{c|}{\textbf{0.0406}}       & \textbf{0.0087}         & \textbf{0.1539}          & \multicolumn{1}{c|}{\textbf{0.0775}}        & \textbf{0.0081}         & \textbf{0.1511}          & \textbf{0.0750}   \\ \hline\hline
\multicolumn{1}{c|}{\textbf{Improvements (\%)}}                   & \textbf{30.46}          & \textbf{30.32}           & \multicolumn{1}{c|}{\textbf{28.75}}       & \textbf{29.86}         & \textbf{30.03}           & \multicolumn{1}{c|}{\textbf{33.67}}       & \textbf{14.65}          & \textbf{13.06}           & \multicolumn{1}{c|}{\textbf{17.90}}        & \textbf{19.26}         & \textbf{17.73}           & \textbf{19.16}  \\ \bottomrule
\end{tabular}
}
\vspace{-0.2cm}
\end{table*}

\vspace{1mm}
\noindent\textbf{Competitors}. To evaluate the effectiveness of \ours, we compared it with seven state-of-the-art competitors. They can be divided into two groups: GCN-based recommender systems (NGCF~\cite{NGCF} and LightGCN~\cite{LightGCN}) and multimedia recommender systems (VBPR~\cite{VBPR}, MMGCN~\cite{MMGCN}, GRCN~\cite{GRCN}, LATTICE~\cite{LATTICE}, and MARIO~\cite{MARIO}). For GCN-based recommender systems, we used user--item interactions {\it only}. For multimedia recommender systems, we used not only user--item interactions but also textual and visual modalities.

\vspace{1mm}
\noindent\textbf{Evaluation Protocols}. For testing, in each dataset, we randomly selected 80\% of the interactions of each user to construct the training set, other 10\% of those for the validation set, and the remaining 10\% of those for the test set, which is the same as in~\cite{MMGCN, GRCN, LATTICE, MARIO, lattice_IEEE}. Then, we performed top-20 recommendations by using each method and then evaluated their accuracies in terms of precision, recall, and normalized discounted cumulative gain (NDCG). We report the average of values obtained by performing the five independent evaluations for each measure.

\vspace{1mm}
\noindent\textbf{Implementation Details}. Following~\cite{LATTICE, MARIO, lattice_IEEE}, for textual and visual modalities, we used the 1,024- and 4,096-dimensional features extracted by the pre-trained sentence-transformers~\cite{SentenceBert} and ImageNet~\cite{ImageNet}, respectively. For a fair comparison, we set the dimensionality of the embedding $d$ as $64$ as in~\cite{MMGCN, GRCN, LATTICE, MARIO, lattice_IEEE} for all methods including \ours. Then, we used the best hyperparameters of competitors and {\ours} obtained by extensive grid search on the validation set in the following ranges: $\{0.0001, 0.0005, 0.001, 0.005, 0.01\}$ for learning rate; $\{0, 0.00001, 0.0001,$ $0.001, 0.01\}$ for regularization weight $\lambda$; $\{1, 2, 3, 4\}$ for the number of GCN-layers $L$ of the methods using GCN.
For \ours, we set its hyperparameters as follows: learning rate = $0.0001$; $\lambda$ = $0.00001$; $L$ = $2$; $\alpha$ = $1$; $\beta$ = $0.5$ for Toys \& Games and $0.3$ for other datasets.

\subsection{Results and Analysis}\label{sec:results}
Due to space limitations, for RQs except for RQ1, we show here only the result on Women Clothing.
Instead, the results on other datasets would be shown at: \url{https://sites.google.com/hanyang.ac.kr/monet-wsdm2024}.
For simplicity, we represent precision@20 and recall@20 as P@20 and R@20, respectively, in the following tables and figures. Also, we highlight the best and the second-best results in each column (\ie, measure) of the following tables in bold and underline, respectively.

\vspace{1mm}
\noindent\textbf{RQ1: Comparison with Seven Competitors}.
Table~\ref{table:sota} shows the results of all competitors and \ours. Our findings are summarized as follows:
\vspace{-2mm}
\begin{longenum}
\item Most importantly, {\ours} {\it consistently} and {\it significantly} outperforms {\it all} competitors on {\it all} datasets for {\it all} measures. Specifically, on Women Clothing, Men Clothing, Beauty, and Toys \& Games, \ours~outperforms the best competitors (\ie, LATTICE and MARIO) by up to 30.32\%, 30.03\%, 13.06\%, and 17.73\% in terms of recall@20. These are {\it dramatic improvements} in the sense that, as reported in the papers of existing state-of-the-art multimedia recommendation methods~\cite{MMGCN, GRCN, LATTICE, MARIO}, they tend to outperform their best competitor by an average of about 9.15\% and a maximum of 17.62\% in terms of recall@20;
\item An MF-based multimedia recommendation method, VBPR, usually shows a {\it lower} accuracy than that of LightGCN using only user--item interactions, despite using multimodal features additionally. This indicates that explicitly capturing collaborative signals through GCN plays an important role in improving the accuracy of multimedia recommendation;
\item MMGCN consistently shows a {\it lower} accuracy than LightGCN, despite using multimodal features additionally. This is because, as shown in Section~\ref{sec:preliminaries}, the non-linear propagation extremely disturbs the reflection of modality features into the final embeddings. Also, the non-linear propagation adversely affects the reflection of collaborative signals~\cite{LightGCN, LRGCCF}. This implies that using linear propagation plays a key role in improving the accuracy;
\item MARIO usually outperforms most of the other competitors because it tries to preserve modality features into the final embeddings via the MP loss. This result once again supports our claim that sufficiently reflecting modality features into the final embeddings is important for improving the accuracy.
\end{longenum}

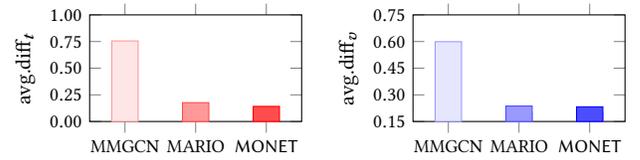
\begin{figure}[t]
\centering
\begin{tikzpicture}
    \small
    \begin{axis}[
        ybar=0pt,
        width=4.4cm,
        height=3.0cm,
        bar width=0.35cm,
        bar shift=0pt,
        ylabel={avg.diff$_t$},
        ymin=0, ymax=1.0,
        symbolic x coords={MMGCN, MARIO, \ours},
        xtick=data,
        ytick={0, 0.25, 0.5, 0.75, 1.0},
        x tick label style={font=\footnotesize, /pgf/number format/.cd,fixed,fixed zerofill,precision=2,/tikz/.cd},
        y tick label style={font=\footnotesize, /pgf/number format/.cd,fixed,fixed zerofill,precision=2,/tikz/.cd},
        enlarge x limits=0.25,
        ]
        \addplot [ybar, draw=red!40, fill=red!10] coordinates {
            (MMGCN, 0.7559)
            (MARIO, 0)
            (\ours, 0)
          };
        \addplot [ybar, draw=red!70, fill=red!40] coordinates {
          (MARIO, 0.1767)};
        \addplot [ybar, draw=red!100, fill=red!70] coordinates {
          (\ours, 0.1417)};
    \end{axis}
\end{tikzpicture}
    \hspace{0.2cm}
\begin{tikzpicture}
    \small
    \begin{axis}[
        ybar=0pt,
        width=4.4cm,
        height=3.0cm,
        bar width=0.35cm,
        bar shift=0pt,
        ylabel={avg.diff$_v$},
        ymin=0.15, ymax=0.75,
        symbolic x coords={MMGCN, MARIO, \ours},
        xtick=data,
        ytick={0.15, 0.3, 0.45, 0.6, 0.75},
        x tick label style={font=\footnotesize, /pgf/number format/.cd,fixed,fixed zerofill,precision=2,/tikz/.cd},
        y tick label style={font=\footnotesize, /pgf/number format/.cd,fixed,fixed zerofill,precision=2,/tikz/.cd},
        enlarge x limits=0.25,
        ]
        \addplot [ybar, draw=blue!40, fill=blue!10] coordinates {
            (MMGCN, 0.5992)
            (MARIO, 0.2)
            (\ours, 0.2)
          };
        \addplot [ybar, draw=blue!70, fill=blue!40] coordinates {
          (MARIO, 0.2372)};
        \addplot [ybar, draw=blue!100, fill=blue!70] coordinates {
          (\ours, 0.2326)};
    \end{axis}
\end{tikzpicture}
\vspace{-0.3cm}
\caption{Average differences for textual and visual modalities of MMGCN, MARIO, and \ours.
}\label{fig:avg_diff}
\vspace{-0.5cm}
\end{figure}

\vspace{1mm}
\noindent\textbf{RQ2: Effectiveness of {\ours} in Reflecting Modality Features}.
To verify that {\ours} better reflects modality features into the final embeddings than other methods (\spec, MMGCN and MARIO)\footnote{Among the existing GCN-based multimedia recommender systems, only MMGCN utilizes multimodal features {\it for the purpose of representing user/item embeddings}. Also, MARIO is the first attempt to preserve modality features into the final item embeddings.}, we compare {\ours} with them in terms of $\text{avg.diff}_m$ for $\forall m\in\{t,v\}$ as in Section~\ref{sec:preliminaries}. Recall that the smaller the score of $\text{avg.diff}_m$, the better the modality $m$'s features are reflected in the final embeddings.

Figure~\ref{fig:avg_diff} shows the results of MMGCN, MARIO, and {\ours}.
For $\forall m\in\{t,v\}$, the scores of avg.diff$_m$ of MARIO are considerably smaller than those of MMGCN. These results are rather obvious because MARIO does not perform neighborhood aggregation via GCN and tries to preserve modality features in the final embeddings. However, note that, for $\forall m\in\{t,v\}$, {\ours} provides comparable avg.diff$_m$ scores to those of MARIO, despite performing neighborhood aggregation via MeGCN. This supports that the design of MeGCN helps reflect modality features into the final embeddings. Furthermore, by using MeGCN, which captures users’ preferences that can be revealed by the {\it modality features of one/multi-hop neighbors}, {\ours} outperforms MARIO dramatically (see Table~\ref{table:sota}).

\begin{table}[t]
\centering
\caption{Accuracies of \ours-MeGCN, \ours-TA, and {\ours}.
All improvements are statistically significant with $p$-value $\leq 0.001$.}\label{fig:megcn_ta}
\vspace{-0.2cm}
\label{table:megcn_ta}
\footnotesize
\resizebox{.36\textwidth}{!}{
\begin{tabular}{c|ccc}
\toprule
\textbf{Methods} & \textbf{P@20} & \textbf{R@20} & \textbf{NDCG@20} \\ \midrule
\textbf{\ours-MeGCN} & 0.0040  & 0.0791  & 0.0361  \\
\textbf{\ours-TA} & \underline{0.0044}  & \underline{0.0866}  & \underline{0.0393}  \\ \hline
\textbf{\ours} & \textbf{0.0050}  & \textbf{0.0990}  & \textbf{0.0450}  \\ \bottomrule
\end{tabular}
}
\vspace{-0.3cm}
\end{table}

\noindent\textbf{RQ3: Effectiveness of MeGCN and Target-Aware Attention}.
{\ours} employs MeGCN and target-aware attention as two core modules.
To verify the effectiveness of these modules, we compare \ours\ with its two variants, namely \ours-MeGCN and \ours-TA, that employ LightGCN {\it instead of} MeGCN and {\it only} MeGCN without target-aware attention, respectively.

As shown in Table~\ref{table:megcn_ta}, \ours~outperforms the two variants for {\it all} measures. Specifically, \ours~outperforms \ours-MeGCN (\textit{resp}. \ours-TA) by up to 24.87\%, 25.08\%, and 24.49\% (\textit{resp}. 14.28\%, 14.26\%, and 14.36\%) in terms of precision/recall/NDCG@20. These results indicate that both MeGCN and target-aware attention play a crucial role in improving the accuracy of multimedia recommendation. Furthermore, it is worth noting that \ours-MeGCN (\textit{resp}. \ours-TA) outperforms {\it all} other state-of-the-art competitors for {\it all} measures (see Tables~\ref{table:sota} and ~\ref{table:megcn_ta}) even though it is {\it not} equipped with MeGCN (\textit{resp}. target-aware attention). Specifically, \ours-MeGCN (\textit{resp}. \ours-TA) outperforms the best competitor (\ie, MARIO) by up to 3.09\%, 4.13\%, and 3.49\% (\textit{resp}. 12.64\%, 14.00\%, and 12.65\%) in terms of precision/recall/NDCG@20. These findings validate that either MeGCN or target-aware attention alone is beneficial in improving the accuracy of multimedia recommendation.

\vspace{1mm}
\noindent\textbf{RQ4: Hyperparameter Sensitivity Analysis}. The parameter $\alpha$ in Eq.~\eqref{eq:3} controls the degree of reflection of a modality feature in the propagation step of MeGCN. We analyze how the accuracy and $\text{avg.diff}_m$ of \ours~change according to different values of $\alpha$ to provide insights into the relationship between collaborative signals and modality features.
From Figure~\ref{fig:alpha}, we observe that the accuracy steadily increases until $\alpha$ reaches 1.0 and then gradually decreases. Specifically, \ours~yields improvements up to 111.91\%, 112.23\%, and 118.73\% in terms of precision/recall/NDCG@20, respectively, when $\alpha$ = $1.0$ compared to the case of $\alpha$ = $0$. We also observe that $\text{avg.diff}_m$ gradually decreases as $\alpha$ increases. Our findings demonstrate that i) small values of $\alpha$ do not sufficiently reflect the {\it modality features} in the final embeddings (\ie, high $\text{avg.diff}_m$ for $\forall m\in\{t,v\}$), resulting in low accuracies; ii) large values of $\alpha$ do not sufficiently reflect {\it collaborative signals} but reflect only modality features in the final embeddings (\ie, low $\text{avg.diff}_m$ for $\forall m\in\{t,v\}$), resulting in low accuracies. These imply that deciding a value of $\alpha$ to {\it properly} reflect {\it both} modality features and collaborative signals plays a crucial role in improving the accuracy.
\definecolor{purple}{rgb}{0.56, 0.0, 1.0}

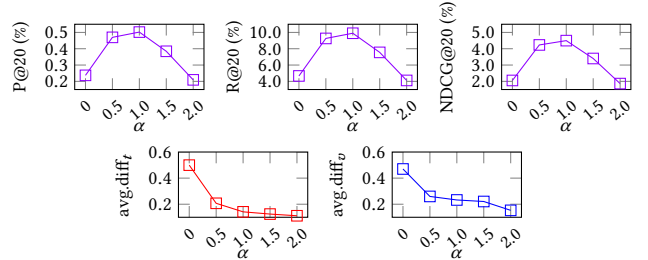
\begin{figure}[t]
\centering
\begin{tikzpicture}
\footnotesize
\begin{axis}[
height=2.45cm,
width=3.3cm,
xtick={1, 2, 3, 4, 5},
xticklabels={0, 0.5, 1.0, 1.5, 2.0},
ylabel=P@20 (\%),xlabel=$\alpha$,
ymin=0.15, ymax=0.55,
xticklabel style={rotate=40, font=\footnotesize},
x label style={at={(0.5,-0.4)}},
y label style={at={(-0.30,0.5)}},
y tick label style={/pgf/number format/.cd,fixed,fixed zerofill,precision=1,/tikz/.cd,font=\footnotesize}]
\addplot[color=purple,mark=square,]
coordinates {(1, 0.236905)(2,0.4697568)(3,0.5020264)(4,0.3843796)(5,0.20926)};
\end{axis}
\end{tikzpicture}
\begin{tikzpicture}
\footnotesize
\begin{axis}[
height=2.45cm,
width=3.3cm,
xtick={1, 2, 3, 4, 5},
xticklabels={0, 0.5, 1.0, 1.5, 2.0},
ylabel=R@20 (\%),xlabel=$\alpha$,
ymin=3.0, ymax=11.0,
xticklabel style={rotate=40, font=\footnotesize},
x label style={at={(0.5,-0.4)}},
y label style={at={(-0.30,0.5)}},
y tick label style={/pgf/number format/.cd,fixed,fixed zerofill,precision=1,/tikz/.cd,font=\footnotesize}]
\addplot[color=purple,mark=square,]
coordinates {(1, 4.6642244) (2, 9.2625926)(3,9.8988256)(4,7.5529862)(5,4.1151702)};
\end{axis}
\end{tikzpicture}
\begin{tikzpicture}
\footnotesize
\begin{axis}[
height=2.45cm,
width=3.3cm,
xtick={1, 2, 3, 4, 5},
xticklabels={0, 0.5, 1.0, 1.5, 2.0},
ylabel=NDCG@20 (\%),xlabel=$\alpha$,
ymin=1.5, ymax=5.5,
xticklabel style={rotate=40, font=\footnotesize},
x label style={at={(0.5,-0.4)}},
y label style={at={(-0.30,0.5)}},
y tick label style={/pgf/number format/.cd,fixed,fixed zerofill,precision=1,/tikz/.cd,font=\footnotesize}
]
\addplot[color=purple,mark=square,]
coordinates {(1, 2.0557044) (2, 4.2320396)(3,4.4963654)(4,3.4015278)(5,1.8639846)};
\end{axis}
\end{tikzpicture}
\begin{tikzpicture}
\footnotesize
\begin{axis}[
height=2.45cm,
width=3.3cm,
xtick={1, 2, 3, 4, 5},
xticklabels={0, 0.5, 1.0, 1.5, 2.0},
ylabel=avg.diff$_t$,xlabel=$\alpha$,
ymin=0.10, ymax=0.60,
xticklabel style={rotate=40, font=\footnotesize},
x label style={at={(0.5,-0.4)}},
y label style={at={(-0.30,0.5)}},
y tick label style={/pgf/number format/.cd,fixed,fixed zerofill,precision=1,/tikz/.cd,font=\footnotesize}]
\addplot[color=red,mark=square,]
coordinates {(1, 0.499489786) (2,0.207062892)(3,0.141669308)(4,0.124600075)(5,0.111582679)};
\end{axis}
\end{tikzpicture}
\begin{tikzpicture}
\footnotesize
\begin{axis}[
height=2.45cm,
width=3.3cm,
xtick={1, 2, 3, 4, 5},
xticklabels={0, 0.5, 1.0, 1.5, 2.0},
ylabel=avg.diff$_v$,xlabel=$\alpha$,
ymin=0.10, ymax=0.60,
xticklabel style={rotate=40, font=\footnotesize},
x label style={at={(0.5,-0.4)}},
y label style={at={(-0.30,0.5)}},
y tick label style={/pgf/number format/.cd,fixed,fixed zerofill,precision=1,/tikz/.cd,font=\footnotesize}]
\addplot[color=blue,mark=square,]
coordinates {(1, 0.47000) (2, 0.259681568)(3,0.232550888)(4,0.220938314)(5,0.153107244)};
\end{axis}
\end{tikzpicture}
\vspace{-0.4cm}
\caption{The effect of $\alpha$ on the accuracies and average differences for textual and visual modalities of \ours.}
\label{fig:alpha}
\vspace{-0.5cm}
\end{figure}
\definecolor{purple}{rgb}{0.56, 0.0, 1.0}

\begin{figure}[t]
\centering
\begin{tikzpicture}
\footnotesize
\begin{axis}[
height=2.45cm,
width=3.3cm,
xtick={1, 2, 3, 4, 5, 6, 7},
xticklabels={0, 0.1, 0.3, 0.5, 0.7, 0.9, 1},
ylabel=P@20 (\%),xlabel=$\beta$,
ymin=0.40, ymax=0.55,
xticklabel style={rotate=50, font=\footnotesize},
x label style={at={(0.5,-0.4)}},
y label style={at={(-0.30,0.5)}},
y tick label style={/pgf/number format/.cd,fixed,fixed zerofill,precision=1,/tikz/.cd,font=\footnotesize}]
\addplot[color=purple,mark=square,]
coordinates {(1, 0.439) (2, 0.475)(3,0.502)(4,0.493)(5,0.467)(6,0.439)(7,0.424)};
\end{axis}
\end{tikzpicture}
\begin{tikzpicture}
\footnotesize
\begin{axis}[
height=2.45cm,
width=3.3cm,
xtick={1, 2, 3, 4, 5, 6, 7},
xticklabels={0, 0.1, 0.3, 0.5, 0.7, 0.9, 1},
ylabel=R@20 (\%),xlabel=$\beta$,
ymin=8.0, ymax=11,
xticklabel style={rotate=50, font=\footnotesize},
x label style={at={(0.5,-0.4)}},
y label style={at={(-0.30,0.5)}},
y tick label style={/pgf/number format/.cd,fixed,fixed zerofill,precision=1,/tikz/.cd,font=\footnotesize}]
\addplot[color=purple,mark=square,]
coordinates {(1, 8.664) (2, 9.366)(3,9.899)(4,9.713)(5,9.198)(6,8.632)(7,8.347)};
\end{axis}
\end{tikzpicture}
\begin{tikzpicture}
\footnotesize
\begin{axis}[
height=2.45cm,
width=3.3cm,
xtick={1, 2, 3, 4, 5, 6, 7},
xticklabels={0, 0.1, 0.3, 0.5, 0.7, 0.9, 1},
ylabel=NDCG@20 (\%),xlabel=$\beta$,
ymin=3.5, ymax=5.0,
xticklabel style={rotate=50, font=\footnotesize},
x label style={at={(0.5,-0.4)}},
y label style={at={(-0.30,0.5)}},
y tick label style={/pgf/number format/.cd,fixed,fixed zerofill,precision=1,/tikz/.cd,font=\footnotesize}]
\addplot[color=purple,mark=square,]
coordinates {(1, 3.932) (2, 4.288)(3,4.496)(4,4.360)(5,4.103)(6,3.841)(7,3.688)};
\end{axis}
\end{tikzpicture}
\vspace{-0.4cm}
\caption{The effect of $\beta$ on the accuracies of \ours.}\label{fig:beta}
\vspace{-0.5cm}
\end{figure}
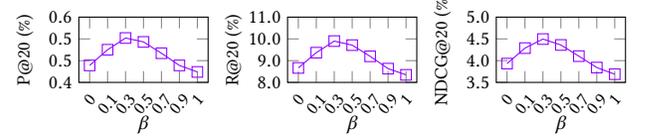

The parameter $\beta$ in Eq.~\eqref{eq:9} adjusts the balance between two matching scores that are computed by leveraging the user's general embedding and target-oriented embedding, respectively. We analyze how the accuracy of \ours~varies according to different values of $\beta$.
As shown in Figure~\ref{fig:beta}, all the accuracies in terms of precision/recall/NDCG@20 steadily increase until $\beta$ reaches $0.3$ and then gradually decrease. Specifically, \ours~yields improvements up to 14.28\%, 14.26\%, and 14.36\% (\textit{resp}. 18.31\%, 18.59\%, and 21.90\%) in terms of precision/recall/NDCG@20, respectively, when $\beta$ = $0.3$ compared to the case of $\beta$ = $0$ (\textit{resp}. $\beta$ = $1$) that utilizes {\it only} general (\textit{resp}. target-oriented) user embedding.
These results indicate that leveraging both general and target-oriented embeddings of each user helps capture the user's preference more precisely for the target item, thus improving the accuracy.
\section{Conclusions and future work}\label{sec:conclusions}

In this paper, we demonstrated the following two important points in the sense of capturing a user's preference for a target item more precisely: (1) {\it both} multimodal features and collaborative signals in user--item interactions should be well reflected in the final user/item embeddings; (2) it is required to generate a {\it target-oriented user embedding} where the interests in her interacted items more relevant to the target item w.r.t. multimodal features are more reflected.
In light of this, we proposed a novel GCN-based multimedia recommendation method, named {\ours}, built upon MeGCN and target-aware attention.
Through extensive experiments with four real-world datasets, we demonstrated that i) {\ours} consistently and dramatically outperforms seven state-of-the-art competitors (up to 30.32\% higher accuracy in terms of recall@20, compared to the best competitor) and ii) even employing one of two core modules outperforms the best competitor, thereby making the final \ours~equipped with the two core modules most accurate in multimedia recommendation.

Our methodology in designing modality-embraced embeddings can be easily and effectively applied in other domains where node features are important (\eg, knowledge graphs). Therefore, as potential avenues of future research, we are interested in employing MeGCN to solve relevant downstream tasks in these domains.

\section*{Acknowledgment}
The work of Sang-Wook Kim was supported by the Institute of Information \& Communications Technology Planning \& Evaluation (IITP) grant funded by the Korea government (MSIT) (No. 2022-0-00352, No. RS-2022-00155586, and No. 2020-0-01373).
The work of Won-Yong Shin was supported by the National Research Foundation of Korea (NRF) grant funded by the Korea government (MSIT) under Grant RS-2023-00220762.

\clearpage
\bibliographystyle{ACM-Reference-Format}
\balance
\bibliography{MONET__WSDM_2024_}

\end{document}